\documentstyle[12pt,aaspp4,eqsecnum,flushrt]{article}

\newcommand{\vecA}{\mbox{\boldmath$A$}}      
\newcommand{\vecW}{\mbox{\boldmath$W$}}

\newcommand{\ABt}{\mbox{$A^{({\rm Bar})}_{\theta}$}}      
      
\newcommand{\vecB}{\mbox{\boldmath$B$}}      
\newcommand{\vecH}{\mbox{\boldmath$H$}}      
\newcommand{\vecJ}{\mbox{\boldmath$J$}}      
\newcommand{\vecM}{\mbox{\boldmath$M$}}      
\newcommand{\vecS}{\mbox{\boldmath$S$}}      
\newcommand{\BBtt}{\mbox{$B^{({\rm Bar})}_{\theta\theta}$}}      
\newcommand{\vecOmega}{\mbox{\boldmath$\Omega$}}      

\newcommand{\gcmMMM}{\mbox{g cm${}^{-3}$}}

\newcommand{\Ts}{\mbox{$T_{\rm s}$}}
\newcommand{\Tg}{\mbox{$T_{\rm g}$}}
\newcommand{\TsTg}{\mbox{$\Ts/\Tg$}}
\newcommand{\bfzh}{\mbox{$\hat{ \mbox{\boldmath$z$}} $}}

\newcommand{\tBar}{\mbox{$t_{\rm Bar}$}}
\newcommand{\tgas}{\mbox{$t_{\rm gas}$}}

\newcommand{\pfjh}{\mbox{$\hat{\mbox{\boldmath$J$}}$}}
\newcommand{\pfth}{\mbox{$\hat{\mbox{\boldmath$\theta$}}$}}
\newcommand{\pfph}{\mbox{$\hat{\mbox{\boldmath$\phi$}}$}}
\newcommand{\pbasis}{\mbox{$\left(\pfjh,\pfth,\pfph\right)$}}

\newcommand{\fTE}{\mbox{$f_{\rm TE}$}}
\newcommand{\fTEv}{\mbox{$\fTE\left(\theta\,|\,J\right)$}}
\newcommand{\fv}{\mbox{$f\left(J,\theta\right)$}}

\newcommand{\Erot}{\mbox{$E_{\rm rot}$}}
\newcommand{\deltB}{\mbox{$\delta_{\rm B}$}}

\newcommand{\sint}{\mbox{$\sin\theta$}}

\newcommand{\vth}{\mbox{$v_{\rm th}$}}
\newcommand{\dvrg}{\mbox{\boldmath$\nabla\cdot$}}
\newcommand{\Jtherm}{\mbox{$J_{\rm therm}$}}
\newcommand{\QXMax}{\mbox{$Q_{X,{\rm Max}}$}}
\newcommand{\QXTE}{\mbox{$Q_{X,{\rm TE}}$}}
\newcommand{\QXinf}{\mbox{$Q_{X,{\rm inf}}$}}
\newcommand{\Jeff}{\mbox{$J_{{\rm eff}}$}}

\newcommand{\JMax}{\mbox{$J_{{\rm Max}}$}}

\begin{document}
\title{Barnett Relaxation in Thermally-Rotating Grains}
\author{A. Lazarian\altaffilmark{1} and W. G. Roberge\altaffilmark{2} }
\altaffiltext{1}{Department of Astrophysical Sciences, Princeton University,
Princeton, NJ 08544}
\altaffiltext{2}{Dept.\ of Physics, Applied Physics, \& Astronomy,
Rensselaer Polytechnic Institute, Troy, NY 12180}

\begin{abstract}
We present an exact formulation of the physics of Barnett
relaxation.
Our formulation is based on a realistic kinetic model of the relaxation
mechanism which includes the alignment of the grain angular momentum
in body coordinates by Barnett dissipation,
disalignment by thermal fluctuations, and 
coupling of the angular momentum to the gas via gas damping.
We solve the Fokker-Planck equation for the measure of
internal alignment using numerical integration
of the equivalent Langevin equation for Brownian rotation.
The accuracy of our results is calibrated by comparing our numerical
solutions with exact analytic results obtained for special cases.
We describe an analytic approximation for the
measure of alignment which fits our numerical results
for cases of practical interest.
\end{abstract}

\setcounter{footnote}{0}

\section{Introduction}

The polarization of  starlight, first discovered by
Hiltner (1949) and Hall (1949), implies that the axes
of dust grains are partially aligned with
the interstellar magnetic field, \vecB.
Models of the Davis-Greenstein (DG) effect and other grain
alignment mechanisms generally predict the alignment of
the grain angular momentum, \vecJ, with respect to \vecB\
(for a recent review, see Roberge 1996).
However, to relate the predictions of these models to observations, one
also requires the ``internal distribution'' of \vecJ\ in
grain body coordinates. The classical papers on DG alignment
(Jones \& Spitzer 1967; Purcell \& Spitzer 1971) assumed
that the internal distribution is determined by
gas damping and other external
processes.\footnote{For example, Jones \& Spitzer (1967) noted that
the component of \vecB\ parallel to the grain angular
velocity can enhance the
alignment of \vecJ\ in grain coordinates.}
However, Purcell (1979, hereafter P79)
realized that frictional processes internal to the grain will
dissipate rotational energy into heat on a timescale
that is generally much shorter than the timescales for external interactions.
Since they conserve angular momentum, these internal dissipation
processes tend to drive the grain toward the state of
minimum energy consistent with a fixed value of \vecJ,
rotation about the principal axis of largest rotational
inertia (henceforth the  ``major axis of inertia'').

Several internal dissipation processes were discussed by
P79, who found that ``Barnett dissipation'' is the
dominant effect for typical interstellar grains.
The mechanism is very elegant:
Dolginov \& Mytrophanov (1976) pointed out that a rotating
grain has a magnetization $\vecM = -\chi\vecOmega/\gamma$
due to the Barnett effect, where $\chi$ is the magnetic
susceptibility, $\gamma$ is the magnetogyric ratio of
the paramagnetic spins,
and \vecOmega\ is the grain angular velocity.
It is well known from theoretical
mechanics (e.g., Landau \& Lifshitz 1976) that \vecOmega\
precesses in body coordinates
unless the rotation axis coincides with a principal
axis of the inertia tensor.
Reasoning by
analogy with ordinary paramagnetic dissipation in a time-dependent
external magnetic field,
P79 concluded
that internal dissipation must also be caused by 
the time-dependent ``Barnett equivalent magnetic field,''
$\vecH_{\rm Be} = -\vecOmega/\gamma$.

Virtually all models of interstellar polarization
developed since 1979 have assumed that Barnett dissipation
aligns \vecJ\ {\it perfectly}\/ with the major axis of inertia.
However, Lazarian (1994, henceforth L94) has pointed out that this is a
poor approximation if the grains rotate with thermal kinetic energies:
thermal fluctuations in the Barnett magnetization will excite
rotation about all 3 of the body axes, preventing perfect
alignment unless the rotation is suprathermal or the grain solid
temperature is zero.

The first model of internal relaxation to include 
thermal fluctuations was presented by L94, who modeled
the kinetics of Barnett relaxation by analogy with
the DG effect.
However, this analogy is not exact because, unlike Barnett
relaxation, the DG effect involves external, nonconservative torques.
This difficulty was ignored for the sake of simplicity
by L94 and we now revisit the problem in order to carry
out an exact analysis.
In \S2 we compare the efficiency of various dissipation mechanisms
in order to delineate the circumstances under
which the Barnett effect is the dominant process.
In \S3 we discuss Barnett alignment in an isolated grain
whose angular momentum has a fixed but arbitrary magnitude, $J$.
The value of $J$ is determined by interactions of the grain
with its environment and in \S4 we formulate a kinetic model
wherein the external interactions are provided by gas-grain collisions
and the evaporation of molecules from the grain surface.
We calculate the appropriate measure of internal alignment
numerically in \S5  and calibrate the accuracy of our
results by considering special cases where exact solutions
can be obtained analytically.
Our results are discussed in \S6 and summarized in \S7.

\section{Internal Dissipation}

Inelastic dissipation and Barnett dissipation were identified in
P79 as the two major causes of internal dissipation.
Inelastic dissipation 
is caused by the oscillating stresses associated with the motion
of the grain angular velocity, \vecOmega, in body coordinates.
It is possible to show that the timescale for inelastic dissipation, $t_{ai}$,
scales as $\Omega^{-3}$ compared to $\Omega^{-2}$ for the Barnett
dissipation time, $t_{\rm Bar}$ (see Eq.~(\ref{2.9})). As a result,
inelastic dissipation is the dominant process for
sufficiently large angular velocities.
However, for sufficiently rigid grains, these
velocities lie in the range of suprathermal rotation;
therefore, we consider only Barnett dissipation in the present
paper, which deals with thermally rotating grains.
The interesting problem
of inelastic dissipation in dust grains will be discussed elsewhere.

As we pointed out above, Barnett dissipation tends to
align \vecJ\ with the major axis of inertia (see \S1).
Henceforth we will model the grains as oblate spheroids with semiaxes
$a$ parallel to \bfzh\ and $b$ (with $b>a$) perpendicular to \bfzh,
where \bfzh\ is the symmetry axis (=the major axis of inertia).
Barnett dissipation decreases the angle $\theta$ between \vecJ\
and \bfzh\ at a rate
\begin{equation}
{d\theta \over dt} = -G_{\rm Bar}\sin\theta\cos\theta
\label{2.6}
\end{equation}
(P79; Roberge, DeGraff, \& Flaherty 1993, henceforth RDGF93),
where
\begin{equation}
G_{\rm Bar}=\frac{VKh^2(h-1)J^2}{\gamma^2 I_z^3},
\label{2.7}
\end{equation}
$V$ is the grain volume, $I_x$ is the inertia for
rotation about an axis perpendicular to \bfzh, $h\equiv I_z/I_x$,
$K$ is related to the imaginary part
of the magnetic susceptibility at frequency $\Omega$ by
\begin{equation}
K \equiv{{\rm Im}[\chi\left(\Omega\right)]\over\Omega}\approx 
1.2\times 10^{-13}\left(\frac{T_s}{15K}\right),
\label{2.8}
\end{equation}
and \Ts\ is the grain solid temperature.
The scaling of $K$ in equation (\ref{2.8}) is appropriate for ordinary
paramagnetic materials (see Draine 1996);
for superparamagnetic grains, $K$ should be increased by an uncertain
factor $\sim 10^5$.
We define the Barnett time, \tBar, to be the value of
$G_{\rm Bar}^{-1}$, so that
\begin{equation}
\tBar(J) = {\gamma^2I_z^3 \over VKh^2(h-1)J^2}.
\label{2.9}
\end{equation}
The numerical value of \tBar\ depends on $J$; a typical
angular momentum for thermal rotation at the gas temperature, \Tg,
is
\begin{equation}
J_{\rm therm}=\sqrt{I_z k T_{\rm g}}
\label{2.3}
\end{equation}
and the corresponding numerical value of \tBar\ is
\begin{equation}
\tBar(\Jtherm) = 4\times 10^6 \rho_{{\rm s}0}^2\,T_{{\rm g}1}^{-1}\,
     K_{-13}^{-1}\,b_{-5}^7\,
     {(a/b)\left(1+a^2/b^2\right)^3\over\left(1-a^2/b^2\right)}
     ~~~{\rm s},
\label{2.10}
\end{equation}
for a grain with radius $10^{-5} b_{-5}$ cm and density
$\rho_{{\rm s}0}$ \gcmMMM.
We may say that eq.~(\ref{2.9}) describes Barnett dissipation 
for an individual grain with fixed angular momentum $J$, while
eq.~(\ref{2.10}) decsribes Barnett dissipation for an ensemble
of grains with rotational temperature \Tg.

\section{Barnett Dissipation in an Isolated Grain}

Barnett dissipation is able to transform rotational energy into heat
because coupling exists between the grain rotational and vibrational
degrees of freedom.
However, the existence of this coupling implies that energy can
also be transferred in the opposite direction, i.e.,
from vibration to rotation (see Jones \& Spitzer 1967).
As a result, fluctuations in the thermal energy will randomly excite
rotation about
all three body axes with energies of order $k\Ts$.
If the rotational energy of the grain is $\gg k\Ts$, then these
thermal fluctuations are negligible\footnote{However, see
Lazarian \& Draine 1997, where it is shown that the effects of
incomplete Barnett alignment are important during ``crossovers'' 
in suprathermally rotating grains.}
and Barnett dissipation aligns \vecJ\ perfectly with \bfzh.
This is the assumption adopted by P79, who
was concerned only with rotation at highly superthermal energies.
However, the grains in molecular clouds are likely to have rotational
energies $\sim k\Ts$.
Consequently, it is of great practical interest to determine the
distribution of orientations established by the combined effects
of Barnett dissipation and thermal fluctuations.

This distribution can be calculated analytically for an isolated grain.
Let \fTEv\ be the distribution function for $\theta$, defined
so that $\fTE\,\sin\theta\,d\theta$ is the probability of finding
\vecJ\ oriented between $\theta$ and $\theta+d\theta$.
Here we regard $\fTE$ to be a function of $\theta$ with $J$ a
fixed but arbitrary parameter; the determination of $J$ by external
interactions is discussed in \S4.
Since the grain is assumed to be an isolated system in
thermodynamic equilubrium, \fTE\ 
can be found, in principle, by maximizing the entropy of the combined
rotation/vibration system.
However, the calculation is drastically simplified by noting
that the number of vibrational degrees of freedom ($\sim$$N$, where $N$
is the number of atoms in the grain) is vastly greater than the
number of rotational degrees of freedom (=$3$).
Thus, the rotational degrees of freedom may be regarded
as a small system immersed in a heat bath at constant temperature, 
\Ts.\footnote{Here we assume for simplicity that \Ts\ is
maintained at a constant value by the various processes
(interactions with the ambient radiation field, gas damping, etc)
by which energy is exchanged between the grain and its environment.
That is, we assume that the radiation field, gas temperature, etc
change on a timescale that is much longer than \tBar.}
It follows that \fTE\ is just the Boltzmann distribution,
\begin{equation}
\fTEv = C\,\exp[-\Erot(\theta)/k\Ts],
\label{3.1}
\end{equation}
where $C$ is a normalization constant and
\begin{equation}
\Erot(\theta)={J^2 \over 2I_z}\,\left[1+(h-1)\sin^2\theta\right]
\label{3.2}
\end{equation}
is the rotational energy.

In a steady state, the linear and circular polarization produced by an
ensemble of spheroidal grains depend only on the measure of
internal alignment,
\begin{equation}
Q_X \equiv \frac{3}{2}\left[\left<\cos^2\theta\right>-\frac{1}{3}\right],
\label{3.3}
\end{equation}
where angle brackets denote the distribution average.
Evaluating the average for \fTE, we find that the measure of
internal alignment for an isolated grain is
\begin{equation}
\QXTE\left(a/b,\Ts,J \right) =
{3 \over 2\xi^2}\, \left[\,
{
\int_0^{\xi}\ t^2\,\exp\left(t^2-\xi^2\right)\,dt
\over
\int_0^{\xi}\ \exp\left(t^2-\xi^2\right)\,dt
}
\,\right]
-\frac{1}{2},
\label{3.4}
\end{equation}
where we have introduced the dimensionless ``suprathermality parameter,'' 
\begin{equation}
\xi^2 \equiv  {\left(h-1\right)J^2 \over 2I_zk\Ts}.
\label{3.5}
\end{equation}
The dependence of \QXTE\ on $\xi$ (Fig.\ 1) has a straightforward
physical interpretation.
For a fixed value of $J$, the difference in rotational energy
between the minimum
$(\theta=0)$ and maximum $(\theta=\pi/2)$ rotational energy ``states''
of a grain is $\Delta E = \left(h-1\right)J^2/2I_z$ [see eq.\ (\ref{3.2})].
Thus, $\xi^2$ is just the ratio of the energy required to
disorient \vecJ\ to the energy added to rotation by a typical thermal
fluctuation.
Weak alignment ($\xi\lesssim 1$) occurs when a typical fluctuation
is large enough to disorient \vecJ\ ($k\Ts \gtrsim \Delta E$),
corresponding to nearly-spherical grains ($h \approx 1$) or thermal
rotation energies ($J^2/2I_z \sim k\Ts$).
Strong alignment ($\xi \gg 1$) occurs only for nonspherical
grains and suprathermal rotation.

\section{Barnett Dissipation with External Interactions}

\subsection{Mathematical Description}

The effects of external interactions can be included by solving the
Fokker-Planck equation,
\begin{equation}
\frac{\partial f}{\partial t} = -\dvrg\vecS,
\label{4.1}
\end{equation}
where $t$ is time, $\fv$ is the joint distribution for
$J$ and $\theta$, and \vecS\ is the probability current.
The latter is given in general by 
\begin{equation}
\vecS = \vecA\,f\,-\,\frac{1}{2}\,\dvrg\left(B\,f\right),
\label{4.2}
\end{equation}
where
\begin{equation}
\vecA \equiv \left<{\Delta \vecJ \over \Delta t}\right>
\label{4.3}
\end{equation}
is the mean torque and
\begin{equation}
B \equiv \left<{\Delta \vecJ \Delta \vecJ \over \Delta t}\right>
\label{4.4}
\end{equation}
is the diffusion tensor.
In principle, \vecA\ and $B$ include
the cumulative effects of every process
that changes \vecJ\ in the body frame.
In this paper, we consider the effects of Barnett dissipation
and gas damping but neglect the effects of other external
interactions such as the Davis-Greenstein mechanism.
As a result, our calculations accurately describe the
internal angular momentum distribution only under conditions
where the gas damping time is much smaller than the timescale for 
other external processes.\footnote{In particular,
we do not attempt to describe magnetic dissipation by ordinary
or superparamagnetic grains.
These effects will be considered in subsequent papers on
Davis-Greenstein alignment (Lazarian 1997a; Roberge \&
Lazarian 1997a), which generalize the treatments of DG alignment
given in Roberge et al.\ (1993) and Lazarian (1995a).}
We assume that the grain center of mass is at rest with respect
to the gas; the generalization of the results presented here
to Gold-type alignment will be considered in subsequent
papers (Lazarian 1997b; Roberge \& Lazarian 1997b).
We also assume that the properties of the grains and their
environment are steady on timescales much longer than the
gas damping and Barnett times, so that only the
steady solution of the Fokker-Planck equation is of interest.

\subsection{Diffusion Coefficients for Barnett Relaxation}

It is simplest to calculate $\vecA^{\rm (Bar)}$ and $B^{\rm (Bar)}$,
the mean torque and diffusion tensor associated with Barnett relaxation,
in a spherical polar coordinate system with basis \pbasis, where the polar
axis is parallel to the grain symmetry axis.
The conservation of $J$ by the Barnett effect and the azimuthal symmetry of
an oblate spheroid together imply that the only nonvanishing components
are \ABt\ and \BBtt.
The former follows immediately from equation
(\ref{2.6}) and is given by
\begin{equation}
\ABt = -\frac{J}{t_{\rm Bar}}\,\sin\theta\cos\theta
\label{4.5}
\end{equation}
where $t_{\rm Bar}$ is defined by eq.~(\ref{2.9}).
The analogous expression for \BBtt\ can be determined by noting that,
in the absence of other processes, Barnett relaxation
must establish the thermodynamic equilibrium
distribution [eq.\ (\ref{3.1})].
Further, the rate of every microscopic
process equals the rate of the time-reversed process
in thermodynamic equilibrium (the ``principle of detailed balance''), so
the probability current must vanish at every point in phase space.
Writing out the components of \vecS\ in polar coordinates
and setting $S_{\theta}=0$, we find that detailed balance requires
\begin{equation}
\sint\,\fTE\,\ABt-\frac{1}{2J}\,{\partial \over \partial\theta}
\left[\sint\,\fTE\,\BBtt\right] =0.
\label{4.6}
\end{equation}
Expression (\ref{4.6}) is a first-order differential equation
for \BBtt. The solution is
\begin{equation}
\BBtt =
\frac{J^2}{t_{\rm Bar}\sin\theta}\left[\exp(\xi^2\sin^2\theta)
\int^{1}_{\sin^2\theta}\sqrt{y}\exp(-\xi^2 y){\rm d}y +
\exp(-\xi^2\cos^2\theta)\right]\ .
\label{4.7}
\end{equation}
In solving equation (\ref{4.6}), we have imposed the
condition\footnote{It worth noting
that condition (\ref{4.8}) coincides with the result for
the diffusion coefficient at a critical point obtained
in Zeldovich (1942) for the problem of phase transitions
of the first order.}
\begin{equation}
\BBtt(\frac{\pi}{2}, \xi)=\frac{J^2}{t_{\rm Bar}\xi},
\label{4.8}
\end{equation}
which insures  that the $\BBtt$ is smooth at $\pi/2$:
\begin{equation}
\lim_{\xi\rightarrow\infty}
\frac{ {\rm d}^2\BBtt(\pi/2,\xi)}{{\rm d}\theta^2} <\infty.
\label{4.9}
\end{equation}
When $J^2\ll I_z k T_{\rm s}$ thermal fluctuations easily change its
direction and this causes $\BBtt \rightarrow \infty$ in agreement
with our results.

\subsection{Diffusion Coefficients for Gas Damping}

The diffusion coefficients for gas damping were calculated by
RDGF93 for oblate spheroids on the assumption
that \vecOmega\ is aligned perfectly with \bfzh\
(i.e., perfect Barnett alignment).
However, the diffusion tensor is independent of \vecOmega\
to first order in $\Omega b /\vth$, where
$v_{\rm th}=\sqrt{2k\Tg/m}$ is the gas thermal speed.
We may therefore adopt the results of RDGF93 for
the diffusion tensor with high accuracy.
The nonzero Cartesian components are
\begin{equation}
B^{\rm (gas)}_{xx}=
\frac{2\sqrt{\pi}}{3}nmb^{4}v_{\rm th}^{3}\Gamma_{\bot}(e)
\left(1+\frac{T_{\rm s}}{T_{\rm g}}\right),
\label{4.10}
\end{equation}
\begin{equation}
B^{\rm (gas)}_{yy}=B^{\rm (gas)}_{xx},
\label{4.11}
\end{equation}
and
\begin{equation}
B^{\rm (gas)}_{zz}=
\frac{2\sqrt{\pi}}{3}nmb^{4}v_{\rm th}^{3}\Gamma_{\|}(e)
\left(1+\frac{T_{\rm s}}{T_{\rm g}}\right),
\label{4.12}
\end{equation}
where $n$ is the gas number density and
the geometrical factors $\Gamma_{\|}$ and $\Gamma_{\perp}$
are dimensionless functions of $a/b$.
They
are given in terms of the eccentricity, $e=\sqrt{1-a^2/b^2}$,
by
\begin{equation}
\Gamma_{\parallel}(e) = 
{3 \over 16} \, \left\{\ 
3+4(1-e^2)g(e)-e^{-2}\left[1-(1-e^2)^2g(e)\right]
\right\}
\label{4.13}
\end{equation}
and
\begin{equation}
\Gamma_{\perp}(e) = 
{3 \over 32} \, \left\{\ 
7-e^2+(1-e^2)^2g(e)+
(1-2e^2)\left[1+e^{-2}\left[1-(1-e^2)^2g(e)\right]\right]\right\},
\label{4.14}
\end{equation}
respectively, where
\begin{equation}
g(e) \equiv {1 \over 2e} \ln\left({1+e \over 1-e}\right).
\label{4.15}
\end{equation}

The mean torque due to gas damping vanishes to zeroth order
in $\Omega$, so we cannot adopt the expressions given by RDGF93.
However, one can relate the mean torque to the diffusion tensor
via the principle of detailed balance\footnote{Strictly speaking,
this principle applies only in thermodynamic equilibrium with
$\Ts=\Tg$. However, it can be shown that the mean torque is
independent of $\TsTg$ to first order in $\Omega b/\vth$
so that, apart from a small error, the thermodynamic equilibrium
result equals the result for $\Ts\ne\Tg$.}
with the result
\begin{equation}
A^{\rm (gas)}_x =
-\frac{4\sqrt{\pi}}{3I_x}nmb^{4}v_{\rm th}\Gamma_{\bot}J_x,
\label{4.16}
\end{equation}
\begin{equation}
A^{\rm (gas)}_y =
-\frac{4\sqrt{\pi}}{3I_x}nmb^{4}v_{\rm th}\Gamma_{\bot}J_y,
\label{4.17}
\end{equation}
and
\begin{equation}
A^{\rm (gas)}_z =
-\frac{4\sqrt{\pi}}{3I_z}nmb^{4}v_{\rm th}\Gamma_{\|}J_z. \nonumber\\ 
\label{4.18}
\end{equation}
If the Barnett alignment is perfect, then $J_x=J_y=0$ 
and the preceding expressions reduce to the components
obtained in RDGF93.
The characteristic timescale for gas damping is
\begin{equation}
\tgas = \frac{3I_z}{4\sqrt{\pi}nmb^{4}v_{\rm th}\Gamma_{\|}},
\label{4.19}
\end{equation}
where
\begin{equation}
\tgas = 7 \times 10^9 \,
\rho_{{\rm s}0}\,
n_{4}^{-1}\,
T_{{\rm g}1}^{-1/2}\,
a_{-5}\,
\Gamma_{\|}\ \ {\rm s}
\label{4.20}
\end{equation}
for a gas of pure H$_2$.

\subsection{Numerical Methods}

We solve the Fokker-Planck equation indirectly by integrating
the equivalent ``Langevin equation for Brownian rotation,''
\begin{equation}
d\vecJ = \vecA\,dt + \sqrt{B}\,d\vecW
\label{4.21}
\end{equation}
(RDGF93), where $\sqrt{B}$ denotes the matrix square
root of $B$ and $d\vecW$ is a vector of statistically
independent, Gaussian random variables with variance $dt$.
To calculate $Q_X$, we integrate equation (\ref{4.21}) numerically
to generate a simulation of $\vecJ(t)$ and equate the
ensemble average of $\cos^2\theta$ in eq.\ (\ref{3.3})
to the time average over the history of the simulation.
The numerical integration was performed using the Euler
algorithm; this scheme approximates the change in $J_i$ during
each time step, $\Delta t$, by
\begin{equation}
\Delta J_i = A_i\,\Delta t + \left(\sqrt{B}\right)_{ij}\,\Delta W_j,
\label{4.22}
\end{equation}
where the $\Delta W_j$ are independent samples from a Gaussian distribution
with variance $\left(\Delta t\right)^2$.
The numerical integration was performed using a ``mixed'' scheme,
where the changes in $J_x$, $J_y$, and $J_z$ due to gas damping were 
computed from the Euler algorithm in Cartesian components and
the change in $\theta$ due to Barnett relaxation was computed
from the Euler algorithm in polar coordinates.

\section{Results}

\subsection{Dimensionless Parameters}

The number of independent parameters in the problem
is greatly reduced by rewriting the Fokker-Planck equation
in terms of dimensionless time and angular momentum
variables.
With time measured in units of \tBar\ and angular momentum
measured in units of \Jtherm, only 3 dimensionless parameters
appear:
\begin{equation}
a/b \equiv \mbox{grain axis ratio},
\label{5.1}
\end{equation}
\begin{equation}
\TsTg \equiv {\mbox{dust solid temperature} \over
            \mbox{gas kinetic temperature}},
\label{5.2}
\end{equation}
and
\begin{equation}
\delta_B \equiv {t_{\rm gas} \over t_{\rm Bar}(J_{\rm therm})}.
\label{5.3}
\end{equation}
Henceforth we will refer to $\delta_B$ as the
``Barnett damping parameter.''
In the model adopted here, where gas damping is the only
external interaction, $Q_X$
is completely determined by specifying the values of $a/b$, \TsTg,
and \deltB.

\subsection{Benchmark Tests}

The accuracy of our numerical method can be
calibrated by considering two special cases where
exact analytic solutions for $Q_X$ exist.
Consider first the case of an isolated grain, for which $Q_X=\QXTE$.
To simulate an isolated grain, we assigned $J/\Jtherm$ as
an initial condition, turned off gas damping in our numerical code,
and integrated the Langevin equation with the timestep
and averaging time set to $\Delta t = 1 \times 10^{-3}$
and $T=2 \times 10^4$ dimensionless units, respectively.
The measures of alignment computed for several cases are plotted as
solid circles in Figure~1.
The rms error in $Q_X$ is $4 \times 10^{-3}$ for the numerical
solutions in Figure~1.

Exact solutions can also be obtained in the ``Maxwellian limit,'' $\deltB = 0$,
where the timescale for Barnett relaxation is infinitely long compared
to the timescale for gas damping.
The Maxwellian distribution for $\theta$ is
\begin{equation}
f_{\rm Max}(\theta) = {h \over 4\pi} \,
              \left(\cos^2\theta+h\sin^2\theta\right)^{-3/2}
\label{5.4}
\end{equation}
(Jones \& Spitzer 1967, but note that we have corrected a minor
typographical error) and the corresponding measure of alignment
is
\begin{equation}
\QXMax =
{3 \over 2 \left(1-h^{-1}\right)}\,
\left[\,1-{1\over \sqrt{h-1}}\,\sin^{-1}\left(1-h^{-1}\right)\,\right]
-
\frac{1}{2}.
\label{5.5}
\end{equation}
In the Maxwellian regime, the partial alignment of \vecJ\ with
\bfzh\ is due entirely to the ``inertial asphericity'' of the grains and,
appropriately, $Q_X$ depends only on $a/b$.\footnote{Recall
that $h=2/\left(1+a^2/b^2\right)$ for a homogeneous, oblate spheroid.}
In Figure~2, we compare the exact solution for \QXMax\ 
(solid curve) with numerical solutions calculated for 
a few values of $a/b$ (filled circles). 
The numerical solutions were obtained by integrating
the Langevin equation as a function of time in units
of \tgas\ (since \tBar\ is irrelevant for this case),
with $\Delta t=1\times 10^{-3}$ and $T=2 \times 10^4$ dimensionless
units, respectively.
The rms error in $Q_X$ for the filled circles in Figure~2
is $2 \times 10^{-3}$.

\subsection{Barnett Relaxation with Gas Damping}

Typical results for realistic calculations including 
gas damping and Barnett relaxation are shown in Figures~3 and 4.
As noted above, the case $\delta_B = 0$ corresponds to the Maxwellian
regime and, appropriately, all of the solutions in Figures~3 and 4
converge to the Maxwellian case as $\delta_B \rightarrow 0$.
As $\delta_B$ increases, Barnett relaxation becomes
increasingly important and $Q_X$ diverges from \QXMax.
Notice that Barnett relaxation always {\it enhances}\/ the alignment,
in the sense that $Q_X > \QXMax$ for $\delta_B > 0$.
For large $\delta_B$, the numerical solution for each
combination of $a/b$ and \TsTg\ 
approaches
a limiting value
\begin{equation}
\QXinf\left(a/b,\TsTg\right) \equiv
\lim_{\delta_B \rightarrow\infty} Q_X\left(a/b,\TsTg,\delta_B\right),
\label{5.6}
\end{equation}
which depends only on the grain shape and dust-to-gas temperature ratio.
This limit represents the maximum possible enhancement of
the alignment (relative to \QXMax) which can be produced by Barnett
relaxation in a thermally-rotating grain.
The qualitative behavior of the results presented in Figures~3 and 4
is consistent with the predictions of L94.
For example, the maximum efficiency of Barnett alignment decreases
as \TsTg\ increases and as the grains become more spherical.

Note that the oscillations which occur for large $\delta_B$ in
Figs.\ 3 and 4 are spurious, that is, they are too large to be
statistical fluctuations.
The oscillations are probably an artifact of the ad hoc
prescription we use to treat the singularities in \BBtt\ at
$\theta=0$ and $\theta=\pi$ [see eq.\ (\ref{4.7})]:
in computing \BBtt, we set $\theta = \theta_{\rm min}$ whenever
the trajectory ``visits'' the interval
$\left(0,\theta_{\rm min}\right)$
and similarly for points in the interval
$\left(\pi-\theta_{\rm min},\pi\right)$.
This procedure is motivated by the fact that the fluctuations
in $\theta$ are proportional to $\sqrt{\BBtt}$
[see eq.\ (\ref{4.21})].
For some value of $\theta_{\rm min}$, the fluctuations for
$\theta<\theta_{\rm min}$ become so large that the grain
is completely disoriented and, roughly speaking, the precise values
of the fluctuations ought to be irrelevant.
After experimenting with different choices, we set
$\theta_{\rm min}=0.1$\,rad for the calculations presented in
Figs.\ 3 and 4.

The fact that the solutions in Figures~3 and 4 approach limiting
values for large $\delta_B$ is a consequence of elementary
thermodynamic considerations.
The case $\delta_B \gg 1$ corresponds to conditions where
$\theta$ evolves rapidly (on a timescale $\sim\tBar$) due to the
effects of Barnett relaxation while $J$ evolves slowly
(on a timescale $\tgas \gg \tBar$) due to gas damping.
The disparity between \tBar\ and \tgas\ means that, insofar as the motion
of $\theta$ is concerned, we may regard $J$ as a slowly-varying
parameter.
Thus, if we compute the measure of alignment by averaging the
motion of $\theta$ over times much longer than \tBar\ but
much shorter than \tgas, the result will be $Q_X \approx \QXTE$,
the approximation becoming exact in the limit $\delta_B \rightarrow \infty$.
We conclude that the asymptotic behavior of the solutions in Figures~3
and 4 occurs because, when $\delta_B \gg 1$, the distribution of $\theta$
is always close to the thermodynamic equilibrium distribution
and the latter is independent of $\delta_B$.
Accordingly, we will refer to the case $\delta_B \gg 1$
as the ``thermodynamic regime.''
Of course, \QXTE\ depends on the value of $J$,
so the value of $Q_X$ computed by the ``short-time averaging'' procedure
described above will change quasi-statically as gas damping slowly
changes the magnitude of the angular momentum.
Thus, it is necessary to compute $Q_X$ by averaging
the motion over many gas damping times, as we have done for the
calculations shown in Figures~3 and 4.

For real interstellar grains, the Barnett time is
typically a few orders of magnitude shorter than
the gas damping time [compare eqs.\ (\ref{2.10})
and (\ref{4.20})] so the solutions of practical
interest in future applications correspond to
$\deltB \sim 10^2$--$10^3$.
Although we are unable to compute solutions for
\deltB\ values this large
(because the Langevin equation becomes too stiff),
Figures~3 and 4 suggest that the solutions are already close to the
limit $\deltB\rightarrow\infty$ when $\deltB=25$.
Accordingly, values of $Q_X$ computed for
$\deltB=25$ should be good approximations to the results
of practical interest.
We have calculated the measure of internal alignment
as a function of $a/b$ and \TsTg\ for the case $\deltB=25$;
the resulting values of $Q_X$ appear as the upper entries in Table~1.
The entries for $\TsTg=1$ provide an additional check on our calculations
because the angular momentum distribution must be Maxwellian when
all of the temperatures in the system are equal.
We have plotted these entries as open circles in Figure~2;
the rms error in $Q_X$ is $6\times 10^{-3}$ for these points.

Because the distribution of $\theta$ is always close
to the thermodynamic equilibrium distribution when
\deltB\ is large, one should be able to approximate
the values of $Q_X$ in Table~1 by an expression of the form
\begin{equation}
Q_X \approx \QXTE\left(a/b,\Ts,\Jeff\right),
\label{5.7}
\end{equation}
where \Jeff\ is some ``effective $J$-value.''
We have computed the effective $J$ value for each entry in Table~1
by interpreting expression (\ref{5.7}) to be the {\it definition}\/
of \Jeff. 
In particular, each $Q_X$ value in the table is given exactly by
\begin{equation}
Q_X = \QXTE\left(a/b,\Ts,\alpha\JMax\right),
\label{5.8}
\end{equation}
where
\begin{equation}
J_{\rm Max} =
\left\{\,
\left[1+{(a/b)^2 \over 2}\right]
\left[1+{T_{\rm s} \over T_{\rm g} }\right]
\,\right\}^{1/2}\,J_{\rm therm}
\label{5.9}
\end{equation}
is the rms value of $J$ for a Maxwellian angular momentum
distribution and $\alpha$ is a dimensionless number.
For each parameter combination, the lower entry in Table~1
gives the corresponding value of $\alpha$.
Figure~5 gives a graphical comparison between selected
results from the table and the ``$\alpha=1$'' approximation
The agreement is generally good, although the $\alpha=1$
approximation systematically underestimates $Q_X$ for points with
$Q_X \gtrsim 0.4$.
This systematic discrepancy is likely due to errors in the numerical
solutions associated with the singularity in \BBtt\ at $\theta=0$
rather than the quality of the approximation.
In view of the various uncertainties in grain properties,
e.g., in the rotational temperature and rotational inertias,
we regard the $\alpha=1$ approximation as a reasonable
method for estimating the degree of Barnett alignment
in real grains when gas damping is the dominant external
interaction.

\section{Discussion}

The study of internal relaxation presented in this paper
includes the physics of Barnett relaxation and gas damping
but does not include the Davis-Greenstein effect or Gold's mechanism.
A quantitative study including the omitted processes  is a 
challenging problem, which we address in subsequent papers in
this series (Lazarian 1997a and Roberge \& Lazarian 1997a 
for the Davis-Greenstein mechanism and Lazarian 1997b
and Roberge \& Lazarian 1997b for Gold's mechanism).
Without going into details, we may summarize
the qualitative effects of incomplete Barnett alignment on
the DG and Gold mechanisms as follows.
Because the DG and Gold mechanisms are both concerned with
the alignment of \vecJ\ with respect to \vecB, the wobbling
of a thermally-rotating grain about its angular momentum (which
occurs when Barnett alignment is imperfect)
will result in a decrease in its polarizing
efficiency relative to the case for perfect Barnett alignment.
According to calculations based on the assumption of
perfect Barnett alignment (Lazarian 1994; Roberge, Hanany, \&
Messinger 1995),
the efficiency of Gold alignment depends on the orientation dependence
of a grain's cross section for collisions with streaming gas
particles, in the sense that the alignment becomes more efficient
as this difference increases.
The precession of the grain axes about \vecJ\ which occurs when
Barnett alignment is imperfect will generally reduce this difference.
Consequently, we anticipate that imperfect Barnett alignment
will reduce the efficiency of DG and Gold alignment
for thermally-rotating grains.
Conversely,
the DG and Gold mechanisms affect the efficiency of Barnett alignment.
Paramagnetic dissipation in the interstellar magnetic field
will decrease the Barnett alignment by reducing the value of $J$.
On the other hand, Gold's mechanism {\it increases}\/ $J$
and therefore increases the efficiency of Barnett alignment.
Strangely, incomplete Barnett relaxation increases the
efficiency of paramagnetic alignment if the grains rotate
suprathermally by increasing the correlation in the directions
of \vecJ\ before and after crossover events
(Lazarian \& Draine 1997).

\section{Summary}

The principal results of this paper are as follows:

\begin{enumerate}

\item
We have presented an exact formulation of the physics
of Barnett relaxation which includes
the effects of thermal fluctuations in the grain material.

\item
We have shown that the measure of internal alignment, $Q_X$,
depends only on 3 dimensionless parameters in the approximation
where the external torques on the grain are dominated
by gas damping. These parameters are the grain axis
ratio, $a/b$, the dust-to-gas temperature ratio, \TsTg,
and a ``Barnett damping parameter,'' \deltB.

\item
We have calculated $Q_X$ for a large number
of parameter combinations.

\item
We have obtained an analytic approximation that
accurately reproduces our numerical solutions in the portion of
parameter space relevant to studies of interstellar polarization.

\item
The results of this paper confirm the conclusion of L94,
that the alignment of \vecJ\ with respect to the
axis of major inertia is incomplete for thermally-rotating grains. 
The incomplete alignment stems from  thermal fluctuations
within the grain material and becomes increasingly significant
as the grain temperature rises.
The rotation of a grain with \vecJ\ parallel to its major axis of
inertia corresponds to the minimum kinetic energy for a given $J$ value.
Thermal fluctuations disorient \vecJ\ because they
transfer energy from the grain solid material to the rotational
degrees of freedom.
In a steady state, energy is transferred back to the grain material
at an equal rate by Barnett dissipation.

\item
It is important to include incomplete Barnett alignment in studies
which relate grain alignment theory to polarimetric data. The 
polarization of electromagnetic radiation
is caused by the preferential alignment of the grain axes. At the same time,
theoretical studies of alignment deal with anisotropies in 
the angular momentum distribution.\footnote{In fact, it is essential to know
the orientation of the grain axes even at this stage, as the alignment (e.g., 
Gold alignment) depends on the axis orientation with respect to 
the magnetic field or to the gaseous flow.}
The distribution
of \vecJ\ in grain coordinates is essential for relating
observations to theory.  

\end{enumerate}

\acknowledgements
We thank Bruce Draine and Lyman Spitzer 
for numerous extremely valuable discussions. We are grateful to our
referee
Peter Martin for detailed comments that greatly improved the paper.
A.L. acknowledges the support of NASA grant NAG5~2858.
W.G.R. acknowledges the support of NASA grant NAGW~3001.

%
%

%
%
\newpage
\centerline{\bf FIGURE CAPTIONS}

\bigskip
\noindent
{\bf Fig.~1} ---
Measure of alignment for an isolated grain with constant angular
momentum, plotted versus the square root of the suprathermality
parameter.
Solid curve: exact solution from equation (\ref{3.4}).
Filled circles: approximate solutions obtained by integrating
the Langevin equation.
The rms error in the numerical solutions is $4 \times 10^{-3}$.

\noindent
{\bf Fig.~2} ---
Measure of alignment in the Maxwellian regime plotted versus
the grain axis ratio.
Solid curve: exact solution from equation (\ref{5.5}).
Filled circles: numerical results for $\delta_B=0$ and arbitrary \TsTg.
Open circles: numerical results for $\delta_B=25$ and $\TsTg=1$.
The rms error in the numerical solutions is
$2 \times 10^{-3}$ for the filled circles and
$6 \times 10^{-3}$ for the open circles.

\bigskip
\noindent
{\bf Fig.~3} ---
Numerical calculations of the measure of alignment for
a highly-flattened grain with $a/b=0.1$ including the effects of
gas damping and Barnett relaxation.
Abcissa: Barnett damping parameter, $\delta_B$ [see eq.\ (\ref{5.3})].
Ordinate: Measure of internal alignment, $Q_X$.
Solutions are presented for $\TsTg=0.1$ (filled circles) and
$\TsTg=0.5$ (open circles).
The dashed line is the exact solution in the Maxwellian regime
and corresponds to $\TsTg=1$.

\bigskip
\noindent
{\bf Fig.~4} ---
Similar to Figure 3 but for a modestly flattened grain with
$a/b=0.5$.

\bigskip
\noindent
{\bf Fig.~5} ---
Graphical comparison between the analytic approximation defined
by expression (\ref{5.8}) with $\alpha=1$
(solid curves) and selected numerical results from Table~1 (symbols).
Results are shown for 
$a/b=0.1$ (top curve, filled circles),
$a/b=0.5$ (middle curve, open circles), and
$a/b=0.9$ (bottom curve, filled squares).

%
%
%
\clearpage
\begin{table}
\begin{center}
\begin{tabular}{cllllllllll}
\multicolumn{11}{c}{\bf Table~1 } \\
\multicolumn{11}{c}{\bf Measure of Internal Alignment
       for {\boldmath$\delta_B=25$}}\\
       &        &         \\     \hline\hline
       & \multicolumn{10}{c}{$T_{\rm s}/T_{\rm g}$} \\ \cline{2-11}
\multicolumn{1}{c}{$a/b$} &
\multicolumn{1}{c}{$0.1$} &
\multicolumn{1}{c}{$0.2$} &
\multicolumn{1}{c}{$0.3$} &
\multicolumn{1}{c}{$0.4$} &
\multicolumn{1}{c}{$0.5$} &
\multicolumn{1}{c}{$0.6$} &
\multicolumn{1}{c}{$0.7$} &
\multicolumn{1}{c}{$0.8$} &
\multicolumn{1}{c}{$0.9$} &
\multicolumn{1}{c}{$1.0$} \\
0.1 & 0.54 & 0.39 & 0.29 & 0.24 & 0.22 & 0.18 & 0.18 & 0.16 & 0.14 & 0.13 \\
    & 0.85 & 0.95 & 0.96 & 0.98 & 1.01 & 0.98 & 1.03 & 1.00 & 0.98 & 0.97 \\
0.3 & 0.49 & 0.35 & 0.25 & 0.22 & 0.18 & 0.17 & 0.15 & 0.14 & 0.14 & 0.13 \\
    & 0.85 & 0.95 & 0.95 & 0.98 & 0.98 & 0.99 & 0.99 & 1.00 & 1.02 & 1.01 \\
0.5 & 0.44 & 0.28 & 0.21 & 0.18 & 0.15 & 0.13 & 0.12 & 0.11 & 0.10 & 0.09 \\
    & 0.90 & 0.97 & 0.98 & 1.04 & 1.02 & 1.00 & 1.00 & 1.01 & 1.00 & 0.96 \\
0.7 & 0.30 & 0.17 & 0.13 & 0.11 & 0.09 & 0.08 & 0.07 & 0.07 & 0.06 & 0.06 \\
    & 0.93 & 0.96 & 1.01 & 1.02 & 1.00 & 0.98 & 0.97 & 1.01 & 0.99 & 0.98 \\
0.9 & 0.11 & 0.06 & 0.04 & 0.03 & 0.03 & 0.02 & 0.02 & 0.02 & 0.01 & 0.01 \\
    & 0.98 & 0.97 & 0.97 & 0.97 & 0.94 & 0.93 & 0.88 & 0.84 & 0.82 & 0.81 \\ \hline
\end{tabular}
\end{center}
\end{table}

%
%
\clearpage

\begin{figure}
\begin{picture}(441,216)
\includegraphics{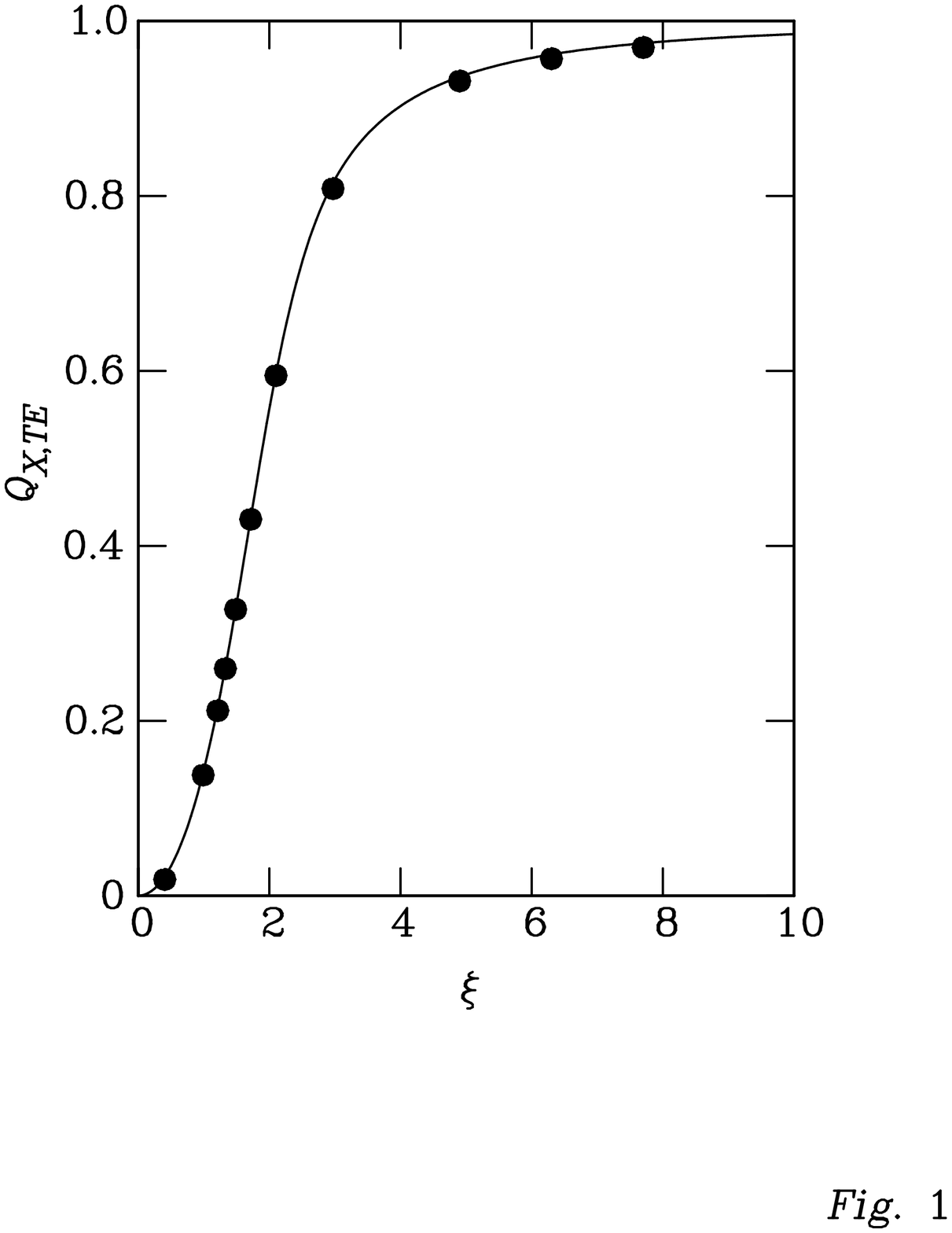}
\end{picture}
\end{figure}

\clearpage

\begin{figure}
\begin{picture}(441,216)
\includegraphics{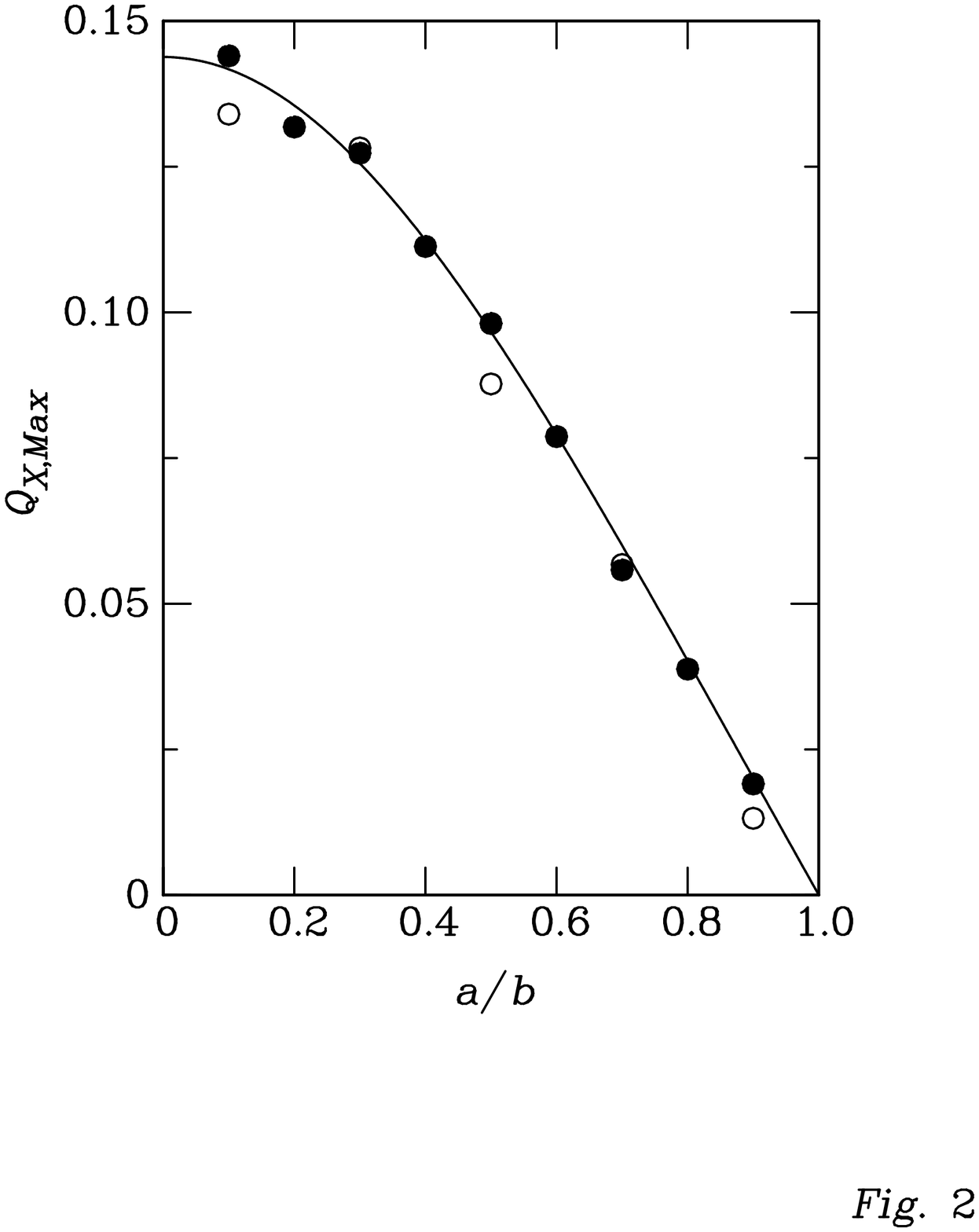}
\end{picture}
\end{figure}

\clearpage

\begin{figure}
\begin{picture}(441,216)
\includegraphics{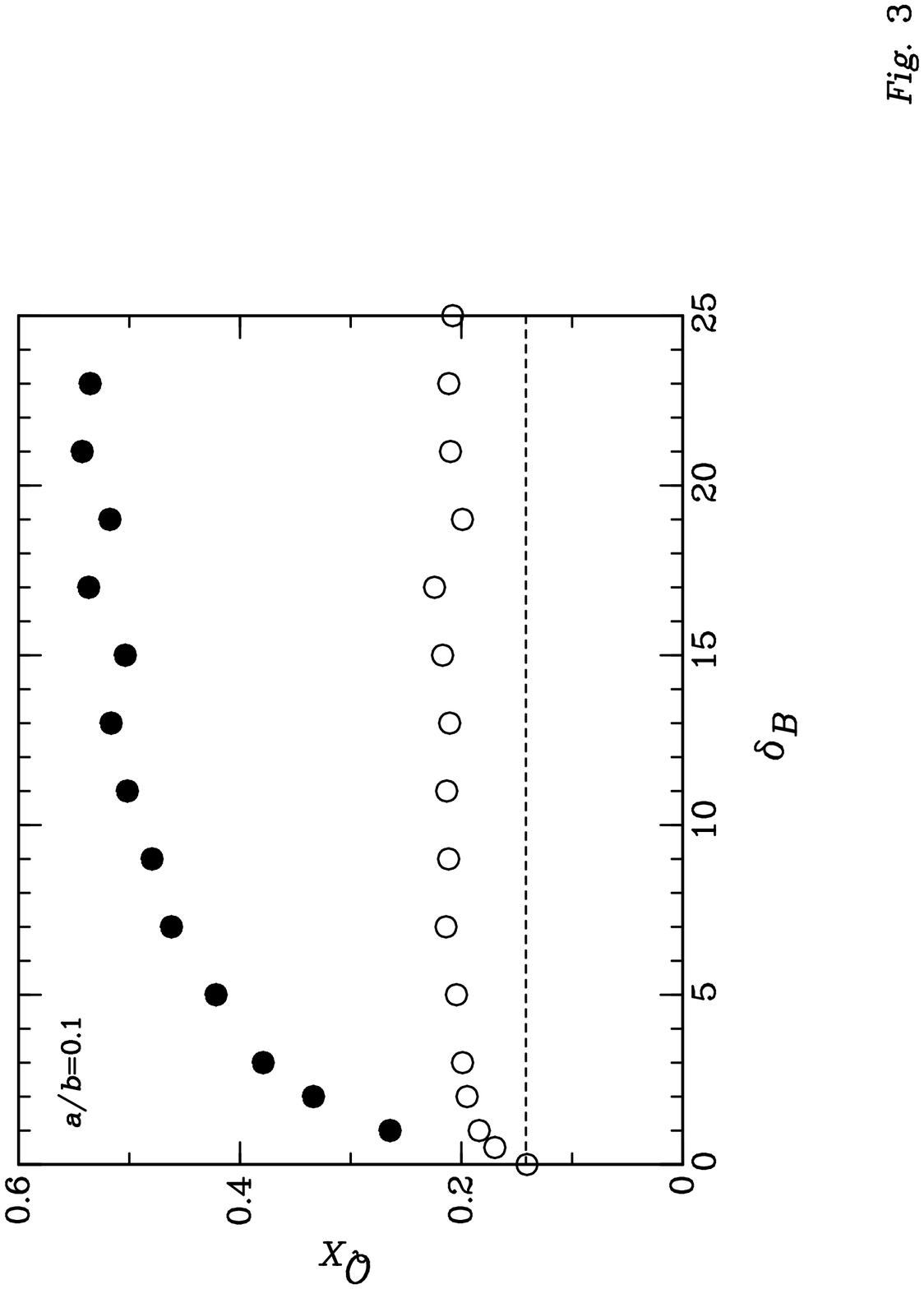}
\end{picture}
\end{figure}

\clearpage

\begin{figure}
\begin{picture}(441,216)
\includegraphics{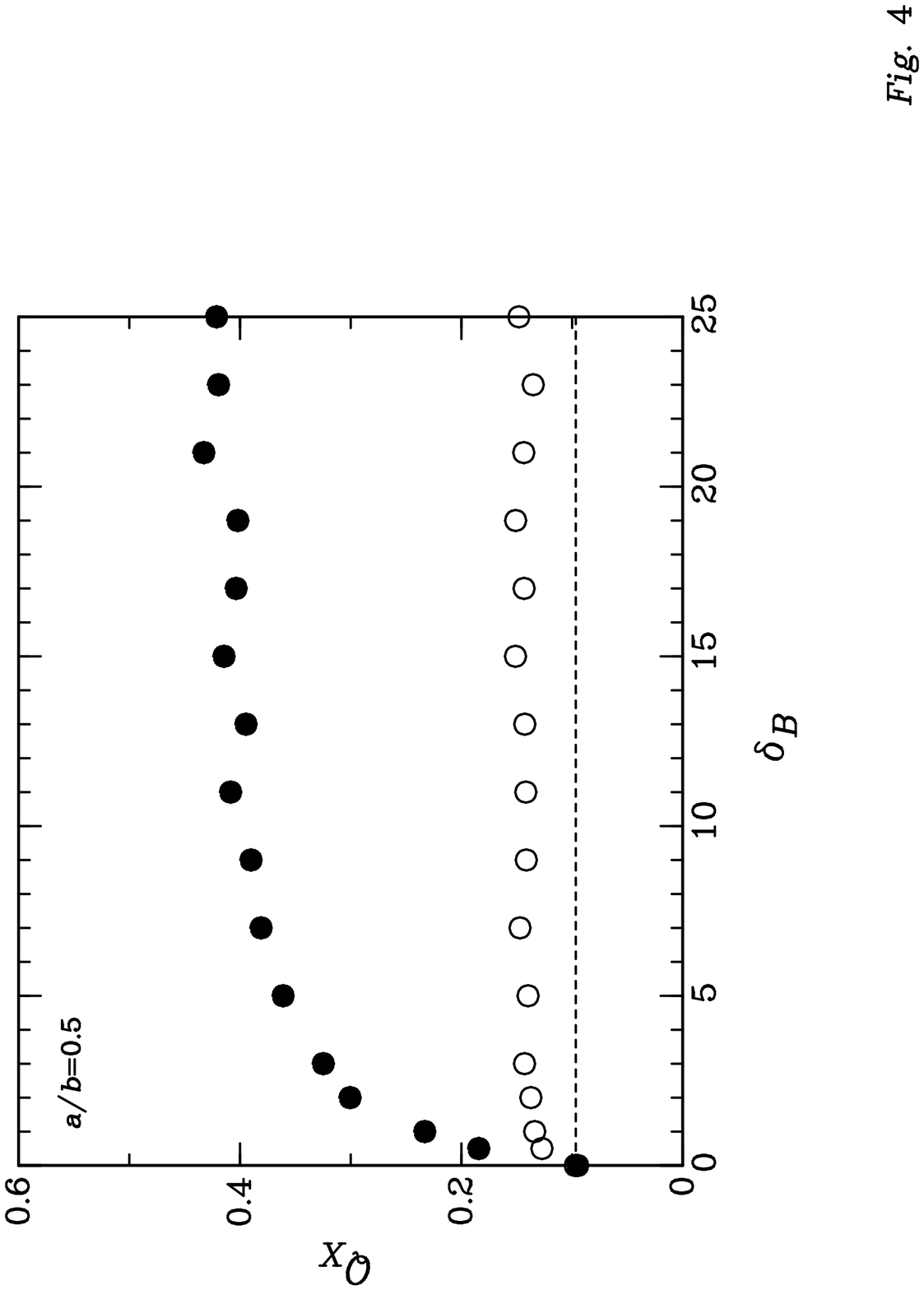}
\end{picture}
\end{figure}

\clearpage

\begin{figure}
\begin{picture}(441,216)
\includegraphics{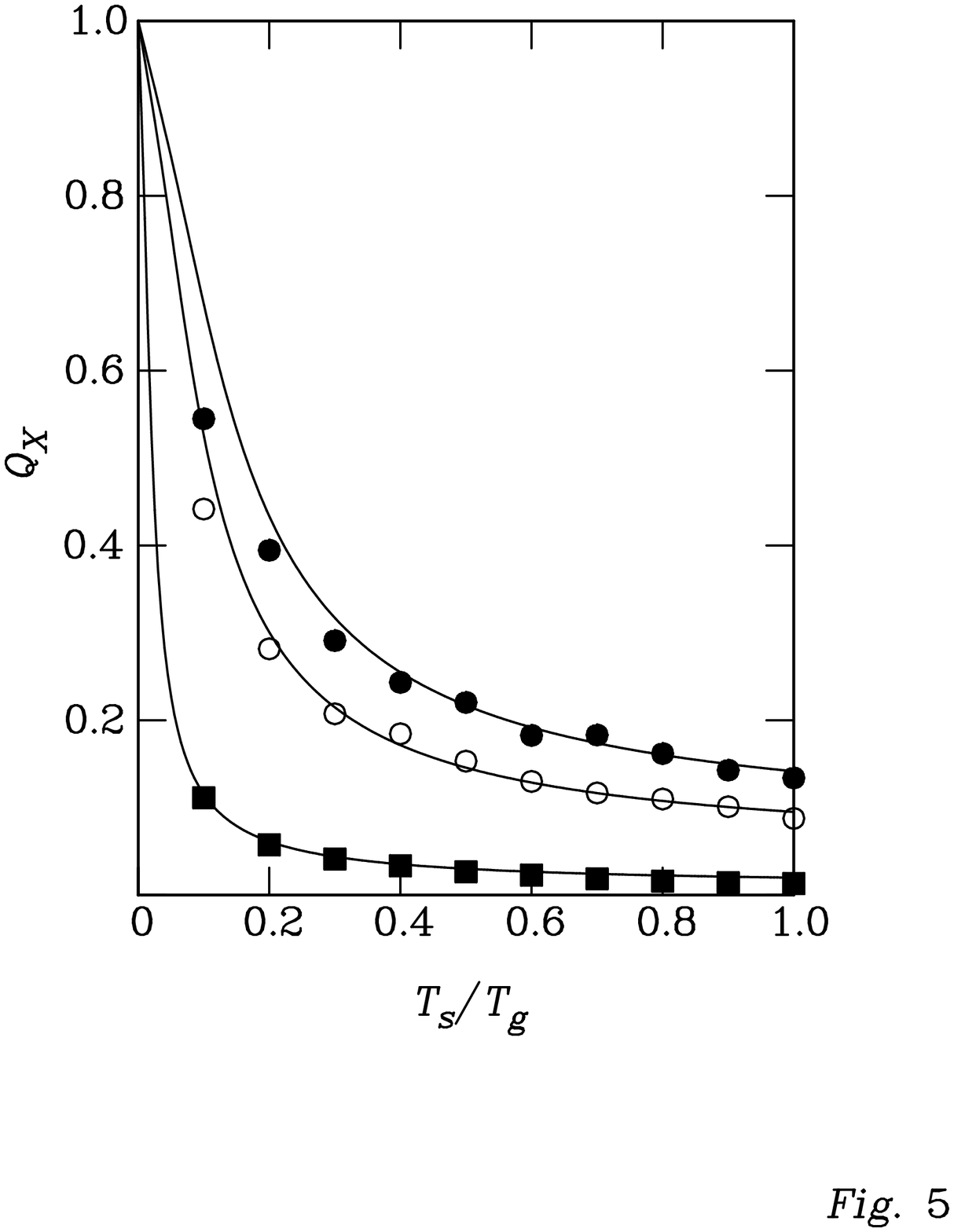}
\end{picture}
\end{figure}

\end{document}